\documentclass[aps,prl,twocolumn,superscriptaddress,longbibliography]{revtex4-2}
\usepackage{amsmath,latexsym}
\usepackage{xcolor}
\usepackage{graphicx}% Include figure files
\usepackage{caption}
\usepackage{color}% color
\usepackage{dcolumn}% Align table columns on decimal point
\usepackage{bm}% bold math
\usepackage{CJK}
\usepackage{braket}
\usepackage{siunitx}
\usepackage[font=small,labelfont=bf]{caption}
\usepackage{hyperref}% add hypertext capabilities
\usepackage[utf8]{inputenc}
\DeclareUnicodeCharacter{2212}{-}
\DeclareUnicodeCharacter{2217}{\ast}

\begin{document}

\title{Van der Waals Spin-Orbit Torque Antiferromagnetic Memory}

\author{Lishu Zhang}\thanks{These authors contributed equally to this work.}
\affiliation{Department of Physics, National University of Singapore, Singapore 117542, Singapore}

\author{Zhengping Yuan}\thanks{These authors contributed equally to this work.}
\affiliation{School of Information Science and Technology, ShanghaiTech University, Shanghai 201210, China}

\author{Jie Yang}
\affiliation{Key Laboratory of Material Physics, School of Physics and Microelectronics, Ministry of Education, Zhengzhou University, Zhengzhou 450001, China}
\author{Jun Zhou}
\affiliation{Institute of Materials Research \& Engineering, A$∗$STAR (Agency for Science, Technology, and Research), Singapore 138634, Singapore}
\author{Yanyan Jiang}
\affiliation{Key Laboratory for Liquid-Solid Structural Evolution and Processing of Materials, Ministry of Education, Shandong University, Jinan 250061, China}
\author{Hui Li}
\affiliation{Key Laboratory for Liquid-Solid Structural Evolution and Processing of Materials, Ministry of Education, Shandong University, Jinan 250061, China}
\author{Yongqing Cai}
\affiliation{Institute of Applied Physics and Materials Engineering, University of Macau, Taipa, Macau SAR, China}

\author{Yuan Ping Feng}
\email{phyfyp@nus.edu.sg}
\affiliation{Department of Physics, National University of Singapore, Singapore 117542, Singapore}
\affiliation{Center for Advanced 2D Materials, National University of Singapore, Singapore 117546, Singapore}
\author{Zhifeng Zhu}
\email{zhuzhf@shanghaitech.edu.cn}
\affiliation{School of Information Science and Technology, ShanghaiTech University, Shanghai 201210, China}
\author{Lei Shen}
\email{shenlei@nus.edu.sg}
\affiliation{Department of Mechanical Engineering, National University of Singapore, Singapore 117542, Singapore}
%\date{\today}

\begin{abstract}
The technique of conventional ferromagnet/heavy-metal spin-orbit torque (SOT) offers significant potential for enhancing the efficiency of magnetic memories. However, it faces fundamental physical limitations, including hunting effects from the metallic layer, broken symmetry for enabling antidamping switching, spin scattering caused by interfacial defects, and sensitivity to stray magnetic fields. To address these issues, we here propose a van der Waals (vdW) field-free SOT antiferromagnetic memory using a vdW bilayer LaBr$_2$ (an antiferromagnet with perpendicular magnetic anisotropy) and a monolayer T$_d$ phase WTe$_2$ (a Weyl semimetal with broken inversion symmetry). By systematically employing density functional theory in conjunction with non-equilibrium Green’s function methods and macrospin simulations, we demonstrate that the proposed vdW SOT devices exhibit remarkably low critical current density approximately 10 MA/cm$^2$ and rapid field-free magnetization switching in 250 ps. This facilitates excellent write performance with extremely low energy consumption. Furthermore, the device shows a significantly low read error rate, as evidenced by a high tunnel magnetoresistance ratio of up to 4250\%. The superior write and read performance originates from the unique strong on-site (insulating phase) and off-site (magnetic phase) Coulomb interactions in electride LaBr$_2$, a large non-zero \textit{z}-component polarization in WTe$_2$, and the proximity effect between them.  
\end{abstract}

\maketitle
%\end{CJK*}
%\tableofcontents

%%\section{Introduction}
\textit{Introduction}---A recent breakthrough in the field of current-induced spin-orbit torque (SOT) produced by ferromagnet/heavy-metal has provided a new and highly efficient method for magnetic state switching without an external magnetic field.\cite{krizakova2022spin, wang2022synergy} However, there are several issues associated with this approach in conventional combination of metallic ferromagnets and bulk heavy metals, such as permalloy/Pt. Firstly, the critical current required for the switching process is relatively high.\cite{kong2020all} Secondly, the switching speed is relatively slow.\cite{krizakova2022spin} Lastly, perpendicular magnetic anisotropy (PMA) MTJ, where the magnetization direction of the magnetic layers is oriented perpendicular to the plane, still necessitates a small in-plane magnetic field.\cite{shukla2022highly, zhang2020rectified} Other critical issues for using 2D ferromagnetic materials include the low Currie temperature, for example 45 K of CrI$_3$,\cite{huang2018electrical} and the hunting effects.\cite{wang2023field}

Recently, Zhou et al., reported that monolayer LaBr$_2$ was a ferromagnetic electride (in which excess electrons act as anions) with both large on-site and off-site Coulomb interactions.\cite{zhou2020atomic} The latter indicates a significant spatial charge fluctuation in LaBr$_2$. In 2022, bilayer LaBr$_2$ was reported to have interlayer antiferromagnetic property and perpendicular magnetic anisotropy.\cite{chen2022electric, sun2022labr2} The large on-site Coulomb interaction leads to stronger exchange interactions between the magnetic moments together with no stray fields in AFM LaBr$_2$, resulting in enhanced stability against thermal or external magnetic perturbations. Meanwhile, the large non-local Coulomb interaction provides a knob for effectively tuning the AFM spin, which is the key issue in conventional AFM spintronics. Thus, bilayer AFM LaBr$_2$ may be a good vdW magnetic material for SOT AFM memories. It is known that a small in-plane magnetic field is required for switching the magnetization of the PMA materials, which is another challenge for achieving ultralow power and miniaturized spintronic devices.\cite{sbiaa2011materials} To realize field-free SOT MRAMs, low symmetric materials have been experimentally used for generating non-zero z-component spin polarization.\cite{dolui2020proximity} Very recently, Kajale et al., experimentally developed all-van der Waals (vdW) SOT devices with ferromagnetic Fe$_3$GaTe$_2$ and low symmetric T$_d$ phase WTe$_2$ for deterministic switching of magnetization of Fe$_3$GaTe$_3$ in the absence of magnetic fields.\cite{kajale2023deterministic}

\begin{figure}[htp!]
\centering
\includegraphics[width=0.5\textwidth]{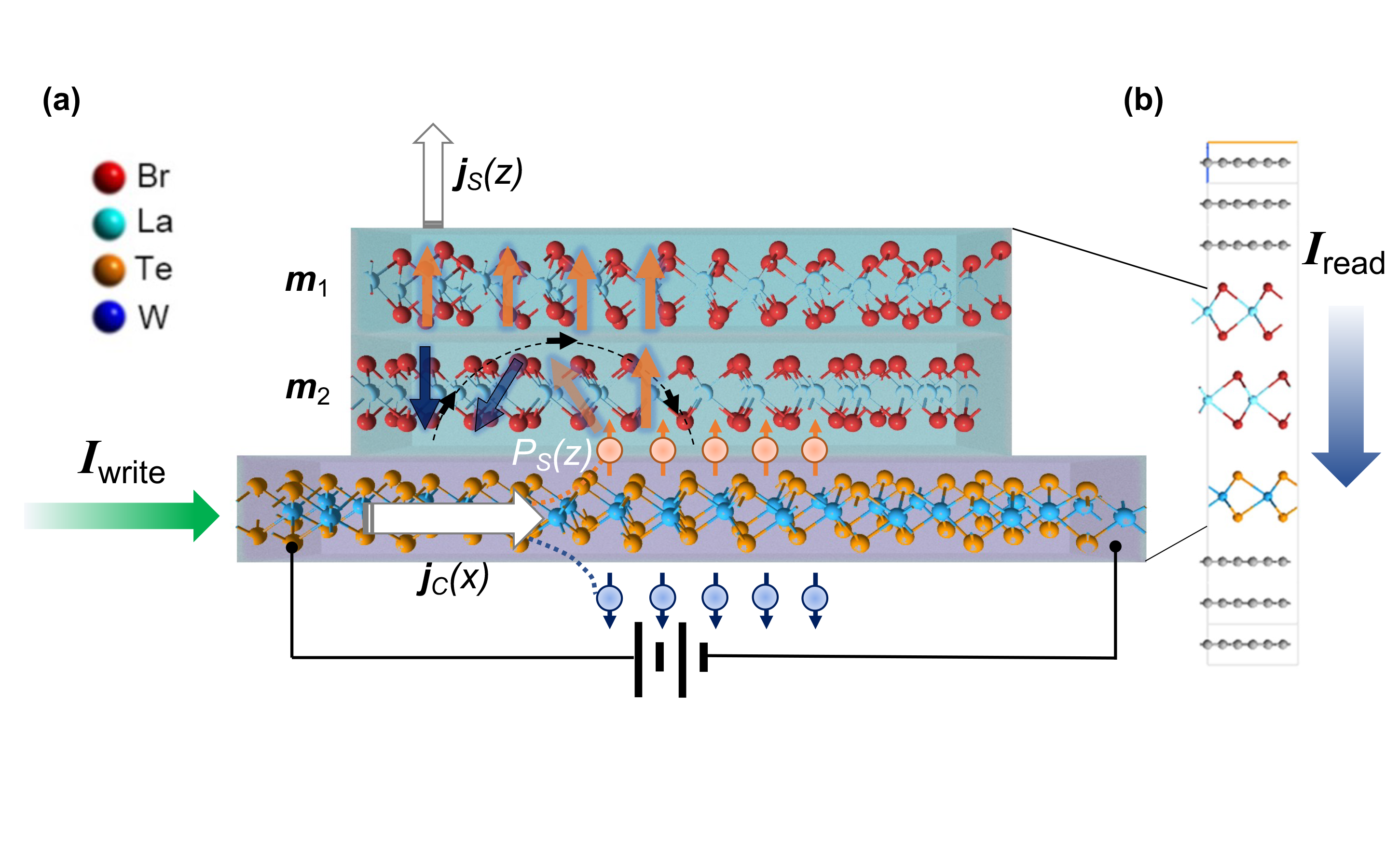}
\caption{Schematic view of the vdW LaBr$_2$/WTe$_2$ SOT antiferromagnetic memory, consisting of a perpendicular antiferromagnetic bilayer LaBr$_2$ and an asymmetric nonmagnetic metallic WTe$_2$. 
(a) \textbf{Write operation:} An in-plane electric current, injected parallel to the interface ($j_c(x)$), generates both an out-of-plane spin current ($j_s(z)$) and spin polarization ($P_s(z)$). The SOT switches the magnetization of the adjacent LaBr$_2$ layer ($\mathbf{m}_2$), while the magnetization $\mathbf{m}_1$ of the top layer remains unchanged. The green arrow indicates the memory writing current. 
(b) \textbf{Read operation:} The TMR effect in the vertical junction is employed to read the magnetic information in the bilayer LaBr$_2$. Two semi-infinite graphene leads facilitate the injection of the read current, indicated by a blue arrow.}
\label{fig:1}
\end{figure}

In this study, we address the challenges in SOT ferromagnetic memories by utilizing 2D vdW antiferromagnetic system composed of bilayer LaBr$_2$ \cite{chen2022electric,sun2022labr2} and monolayer T$_d$ WTe$_2$.\cite{kang2019nonlinear, zhou2019intrinsic} The reason we use T$_d$-phase WTe$_2$ as the SOT source for PMA LaBr$_2$ is that T$_d$-phase WTe$_2$ has a very small lattice mismatch with LaBr$_2$ (1.42 \%). Furthermore, T$_d$-phase WTe$_2$ has broken inversion symmetry and substantial out-of-plane spin Hall conductivity (SHC) \cite{wu2018observation, zhou2019intrinsic, tang2017quantum} which is anisotropic, providing a large polarization in the out-of-plane direction and field-free operation.\cite{van2016field, song2023field, liu2019field} Our results show that this all-vdW field-free SOT antiferromagnetic random-access memory, integrating the write and read functions, has low power consumption through reduced switching current requirements, high-speed magnetic switching in picosecond, and a large tunnel magnetoresistance (TMR) for a low read error rate. Through both analytic analysis and macrospin simulations, we have observed an intrinsic critical current (I$_c$) of $\sim$ 10 MA/cm$^2$ and a switching time (t) of $\sim$ 250 ps in the bilayer LaBr$_2$/WTe$_2$ vdW structure. Regarding the read performance, we have discovered an exceptional tunnel magnetoresistance (TMR) value of 4250\% employing non-equilibrium Green’s function (NEGF) calculations. The outperform write and read performance demonstrates the effectiveness and potential of our proposed 2D vdW AFM LaBr$_2$/WTe$_2$ system for SOT-MRAM applications.

\begin{figure}[htp!]
\centering
\includegraphics[width=0.5\textwidth]{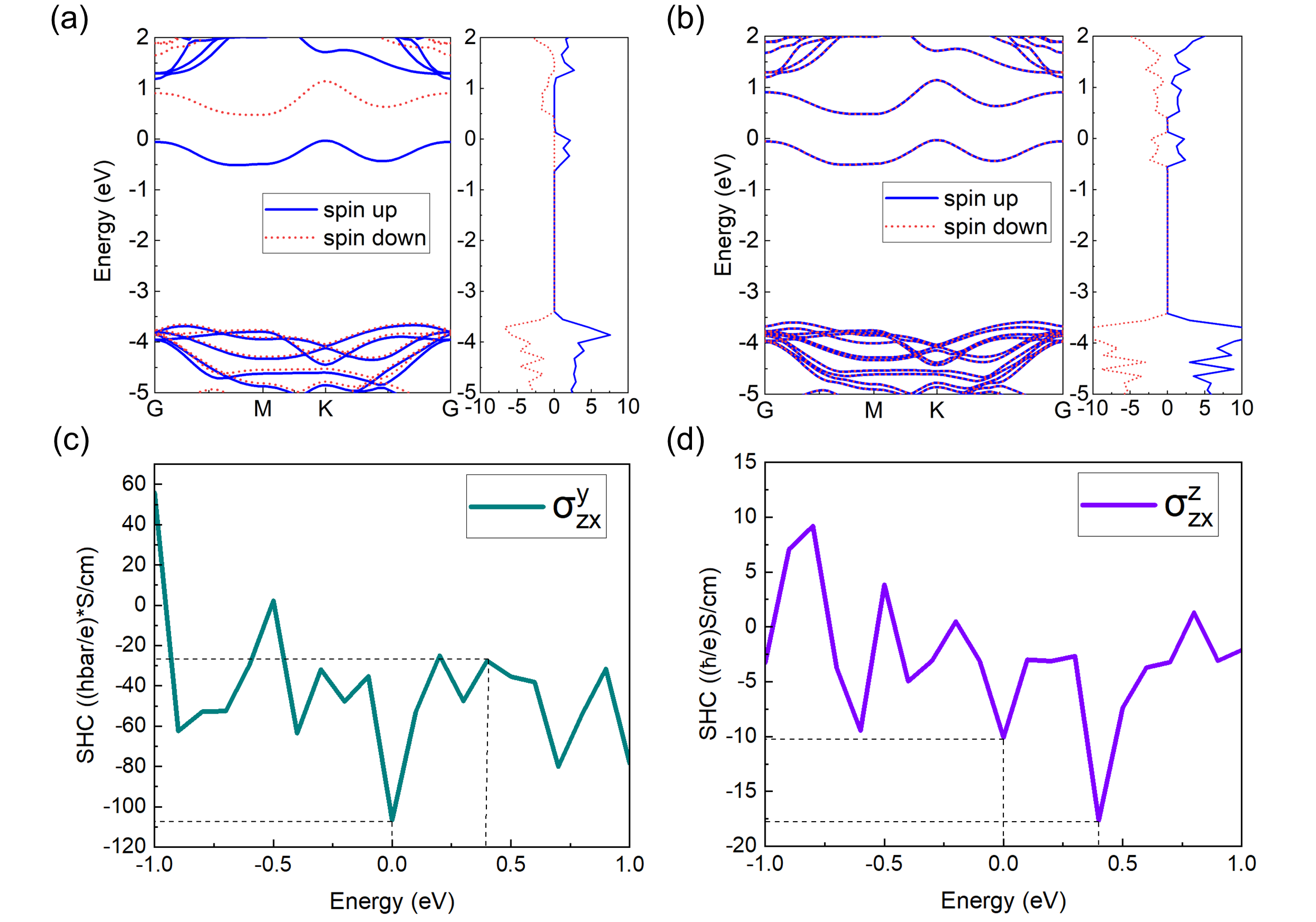}
\caption{The electronic structures of (a) the unit monolayer Labr$_2$, and (b) the unit bilayer LaBr$_2$. (c-d) The variation of SHC with respect to the position of the Fermi energy for monolayer T$_d$ WIe$_2$. The SHC tensor element \( \sigma^{y}_{zx} \), is not equal to zero because of the broken symmetry in T$_d$ WTe$_2$. The Fermi energy is set to zero, at which \( \sigma^{y}_{zx} \)=\textrm{--}106 (h/e)S/cm, \( \sigma^{z}_{zx} \)=\textrm{--}10.5 (h/e)S/cm, and the ratio of \( \sigma^{z}_{zx} \), to \( \sigma^{y}_{zx} \), defined as $tan\beta$, is 0.10.  The ratio will significantly increase to 0.65 when the Fermi level is shifted up to 0.4 eV by electric fields, providing a large out-of-plane spin-polarized current and field-free switching of magnetization of the adjacent LaBr$_2$ layer.}
\label{fig:2}
\end{figure}

\textit{Model}---The atomic schematic diagram of the proposed vdW LaBr$_2$/WTe$_2$-based devices is shown in Fig. \ref{fig:1}. The optimized interlayer distance is indicated in Fig. S1. In Fig. \ref{fig:1}(a), when the write current is injected into the WTe$_2$ substrate, the spin current accumulates at the LaBr$_2$/WTe$_2$ interface and diffuses into the adjacent LaBr$_2$ layer. Consequently, spin-polarized electrons exert a torque on the local magnetic moment by transferring their angular momentum, facilitating the switching of the magnetization $\mathbf{m}_2$ of the LaBr$_2$ layer adjoining the WTe$_2$ layer (as illustrated in Fig. S2). The T$_d$-phase of WTe$_2$, belonging to the space group Pmn2$_1$, exhibits strong SHC and high spin Hall angles (SHA). The band structure of \textit{monolayer} LaBr$_2$ is presented in Fig. \ref{fig:2}(a). The presence of two localized states near the Fermi level corresponds to the spatial electride states, with one state fully occupied and the other fully empty. Figure \ref{fig:2}b shows the \textit{bilayer} LaBr$_2$ which exhibits antiferromagnetic behavior. Further SOC calculations indicate the out-of-plane spin polarization in bilayer LaBr$_2$ ($\mathbf{m}_1$ and $\mathbf{m}_2$ in Fig. \ref{fig:1}).  Two components of the spin Hall conductivity ($\sigma^y_{zx}$ and $\sigma^z_{zx}$) of T$_d$-phase WTe$_2$ are shown in Fig. \ref{fig:2}(c-d). Notably, the $\sigma^z_{zx}$ is one order of magnitude smaller than $\sigma^z_{zx}$ at the Fermi level, while the value of $\sigma^z_{zx}$ is significant at 0.4 eV below E$_f$ and even comparable to the $\sigma^z_{zx}$. This large z-component SHC is in agreement with the experimental value reported for other Weyl semimetals \cite{zhao2020observation}. In the late part, using macromagnetic simulation we will demonstrate how such large ratio of $\sigma^z_{zx}$ over $\sigma^y_{zx}$ can effectively and quickly switch magnetization of $\mathbf{m}_2$ layer of LaBr$_2$ with only small charge current.

\begin{table*} 
  \caption{The DFT and experimental parameters for macromagnetic simulations.}
  \label{table:1}
  \begin{ruledtabular}
  \begin{tabular}{cccc}
    Parameter & Symbol & Value & Remark \\
    \hline
    Magnetic moment per unit cell & $\mu$ & $1 \, \mu \text{B}$ & \\
    Volume of unit cell & $V$ & $8.26 \times 10^{-29} \, \text{m}^3$ & \\
    Saturation magnetization & $M_S$ & $112 \, \text{emu/cm}^3$ & \\
    Magnetic anisotropy energy & $EMA$ & $0.27 \, \text{meV}$ & Out-of-plane \\
    Interlayer exchange & $J_{12}$ & $0.132 \, \text{meV}$ & \\
    Intralayer exchange & $J_{13}$ & $6.38 \, \text{meV}$ & AFM \\
    Magnetic anisotropy constant & $K_u$ & $5.24 \times 10^{5} \, \text{J m}^{-3}$ & \\
    Anisotropic field & $H_k$ & $9.33 \, \text{T}$ & \\
    Exchange field & $H_{ex}$ & $2.27 \, \text{T}$ & \\
    %%Free layer thickness & $t$ & $0.59 \, \text{nm}$ & \\
    Damping constant & $\sigma$ & $0.015$ & \cite{lattery2018low} \\
    Spin Hall angle & $\theta_{SH}$ & $0.323$ & \cite{kang2019nonlinear, zhou2019intrinsic} \\
    %%Spin polarization direction & $\sigma$ & $(0, \cos \beta, -\sin \beta)$ & \\
    Spin Hall conductivity (y-polarization) & $\sigma_{zx}^y$ & $-18$ & At $E_f = - 0.4 \, \text{eV}$ \\
    Spin Hall conductivity (z-polarization) & $\sigma_{zx}^z$ & $-16$ & At $E_f = - 0.4 \, \text{eV}$ \\
    Conductivity angle & $\tan \beta$ & $0.89$ & $(\sigma_{zx}^z) / (\sigma_{zx}^y)$ \\
  \end{tabular}
  \end{ruledtabular}
\end{table*}

\begin{figure*}[htbp!]
\centering
\includegraphics[width=0.9\linewidth]{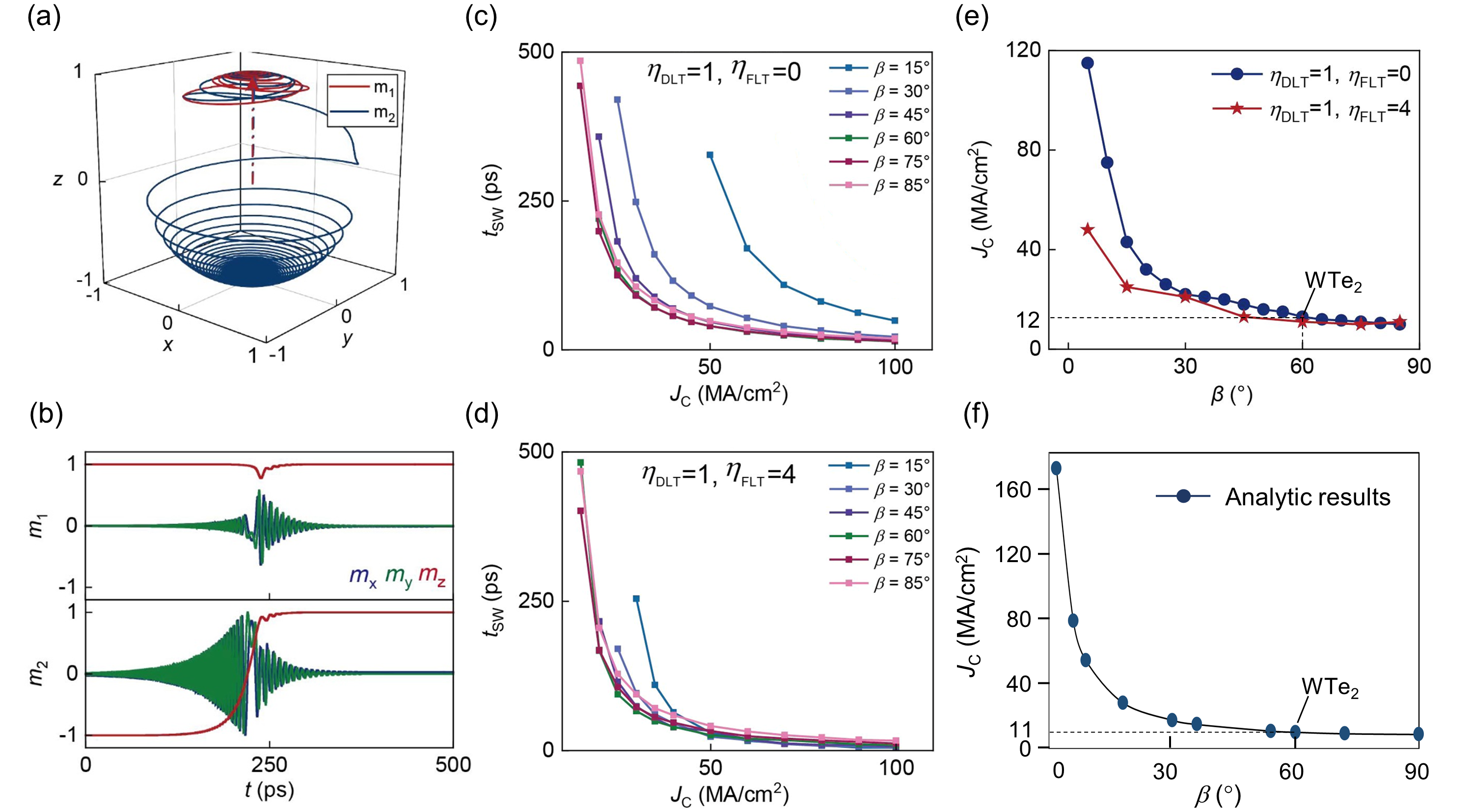}
\caption{The time evolution of (a) the trajectories of magnetic moments and (b) the components of \textbf{m} in the x, y, and z directions. The switching time as a function of applied current density at different spin polarization directions with (c) $\eta_\textrm{DLT}$ = 1 and $\eta_\textrm{FLT}$ = 0 and (d) $\eta_\textrm{DLT}$ = 1 and $\eta_\textrm{FLT}$ = 4. (e) The critical current density as a function of $\beta$. (f) The analytical critical current density is a function of varying $\beta$ when the FLT is zero.}
\label{fig:3}
\end{figure*}

\textit{Write}---To gain insights into the magnetization dynamics of LaBr$_2$, we conducted numerical simulations using the two-sublattice macrospin model (see METHODS for more details), applying a charge current ($\mathbf{J}_{SOT}$) along the $x$-direction to the heavy metal (HM) layer. The device structure is shown in Fig.~\ref{fig:1}, where $\bm{m}_1$ denotes the magnetic moment of the upper LaBr$_2$ layer and $\bm{m}_2$ represents the magnetic moment of the lower layer LaBr$_2$. Meanwhile, the parameters used in the macrospin simulation are summarized in Table~\ref{table:1}, where the spin polarization $\sigma = (0, \cos\beta, -\sin\beta)$. The dynamics of the bilayer LaBr$_2$/WTe$_2$ heterojunction with different spin-$z$ components corresponds to varying $\beta$. Specifically, we chose $\beta = 60\si{\degree}$, corresponding to WTe$_2$ with $E-E_f = 0.4~\text{eV}$, to investigate the switching properties of $\bm{m}_1$ and $\bm{m}_2$. The switching paths of the two magnetic moments are shown in Fig.~\ref{fig:3}(a). Driven by damping-like torque (DLT), the magnetic moment $\bm{m}_2$ undergoes a switching process from the $-z$ direction to the $+z$ direction, while the magnetic moment $\bm{m}_1$ remains in the $−y$ direction. Consequently, $\bm{m}_1$ and $\bm{m}_2$ transition from an antiparallel (AP) state to a parallel (P) state, resulting in the conversion of the system from an AFM mode to an FM mode. Meanwhile, the switching of $\bm{m}_2$ is deterministic and stable. It should be noted that this switching process does not necessitate an external magnetic field and can be accomplished through the application of a DC current, eliminating the need for a two-step process of conventional SOT-driven switching. In addition, the switching process is within 250~ps, as shown in Fig.~\ref{fig:3}(b), which is much shorter than the traditional three-terminal FM device driven by SOT.

%%\subsection{Write operation and performance}

In order to further reveal the impact of $z-$direction spin polarization on SOT switching efficiency, the relationship between switching time ($t_{sw}$) and applied current density $\bm{J}_{\text{SOT}}$ under different $\beta$ in the case of only DLT is investigated. Here, $t_{sw}$ is defined as the time required for the magnetic moment $\bm{m}_2$ to proceed from the initial position $(0, 0, −1)$ to $\bm{m}_{2,z} = 0.9$ \cite{tsou2021thermally}. As shown in Fig. \ref{fig:3}(c), increasing the current density speeds up the switching process for all $\beta$ values. Additionally, Under the same current density, the $t_{sw}$ corresponding to different $\beta$ is significantly different, and a large angle corresponds to faster switching. Furthermore, the impact of field-like torque (FLT) on the system and the coexistence of DLT and FLT are also studied. Similar to the case with only DLT, the magnetic moment $\bm{m}_2$ undergoes switching, and the presence of FLT affects the precession process. Comparing Fig.\ref{fig:3}(b) and Fig. S2(c), it is demonstrated that when the value of FLT is large ($\eta_{FLT} = 4$), its influence on the precession process becomes more pronounced, resulting in a further reduction in the time required for $\bm{m}_2$ switching and a significant decrease in the oscillation amplitude of $m_1$ along the $z$ direction. The presence of FLT facilitates the switching of the magnetic moment $\bm{m}_2$. On the other hand, Fig. \ref{fig:3}(d) reveals that with only FLT, the magnetic moment $\bm{m}_1$ experiences minimal movement, while $\bm{m}_2$ exhibits slight oscillations along the $−z$ direction, even at a high current density ($\bm{J}_{\text{SOT}} = 100$ MA/cm$^2$). This indicates that the switching process of 2D LaBr$_2$ with in-plane anisotropy and spin-$z$ is predominantly governed by DLT, and the presence of FLT influences the precession process. Additionally, we studied the relationship between $t_{sw}$ and $\bm{J}_{\text{SOT}}$ under different $\beta$ under the coexistence of DLT and FLT. As shown in Fig. \ref{fig:3}(d), similar to the case of only DLT, the change of $\beta$ has a significant impact on $t_{sw}$, but it is not as severe as the case of only DLT. At the same time, under the same $\beta$ and $\bm{J}_{\text{SOT}}$, the system where DLT and FLT coexist switches faster, which further proves that FLT promotes switching.

Fig. \ref{fig:3}(e) shows the relationship between critical current density ($\mathbf{J}_c$) corresponding to different $\beta$ measured through simulation. For the two cases where only DLT or DLT is accompanied by FLT, the critical current density decreases significantly with the increase of the presence of a significant out-of-plane SHC, and tends to be stable at larger angles. Conversely, if the material lacks out-of-plane SHC, a significantly higher $\mathbf{J}_c$ would be required. This further emphasizes the significance of selecting T$_d$-pahse WTe$_2$ for our study, rather than other highly symmetric SOC materials.\cite{dolui2020proximity} Moreover, for the case where DLT and FLT coexist, with the same $\beta$, the critical current density is lower than in the case of only DLT. The analytic results of the relationship between $\mathbf{J}_c$ and $\beta$ are shown in Fig. \ref{fig:3}(f), which demonstrates the similarity between numerical simulation and analytic results. A critical current of 12 MA/cm$_2$ for WTe$_2$ in Fig. \ref{fig:3}(e) is similar to 11 MA/cm$^2$ in Fig. \ref{fig:3}(f).

\textit{Read}---Apart from the aforementioned write operation, the read process, which relies on the TMR effect, serves as another crucial function in magnetic memories. To investigate the TMR effect of our vdW LaBr$_2$/WTe$_2$ heterojunction, a vertical model is constructed as shown in Fig. \ref{fig:1}(b). In this model, the P or AP state is defined as the parallel (antiparallel) magnetic orientation between two LaBr$_2$ layers (Fig. \ref{fig:1}(a)). The lower LaBr$_2$ layer serves as the free layer, while the upper LaBr$_2$ layer acts as the pinning layer. A vertical current is injected between semi-infinite graphite leads, sandwiching the LaBr$_2$/WTe$_2$ structure. 
\begin{figure*}[htbp!]
\centering
\includegraphics[width=0.9\textwidth]{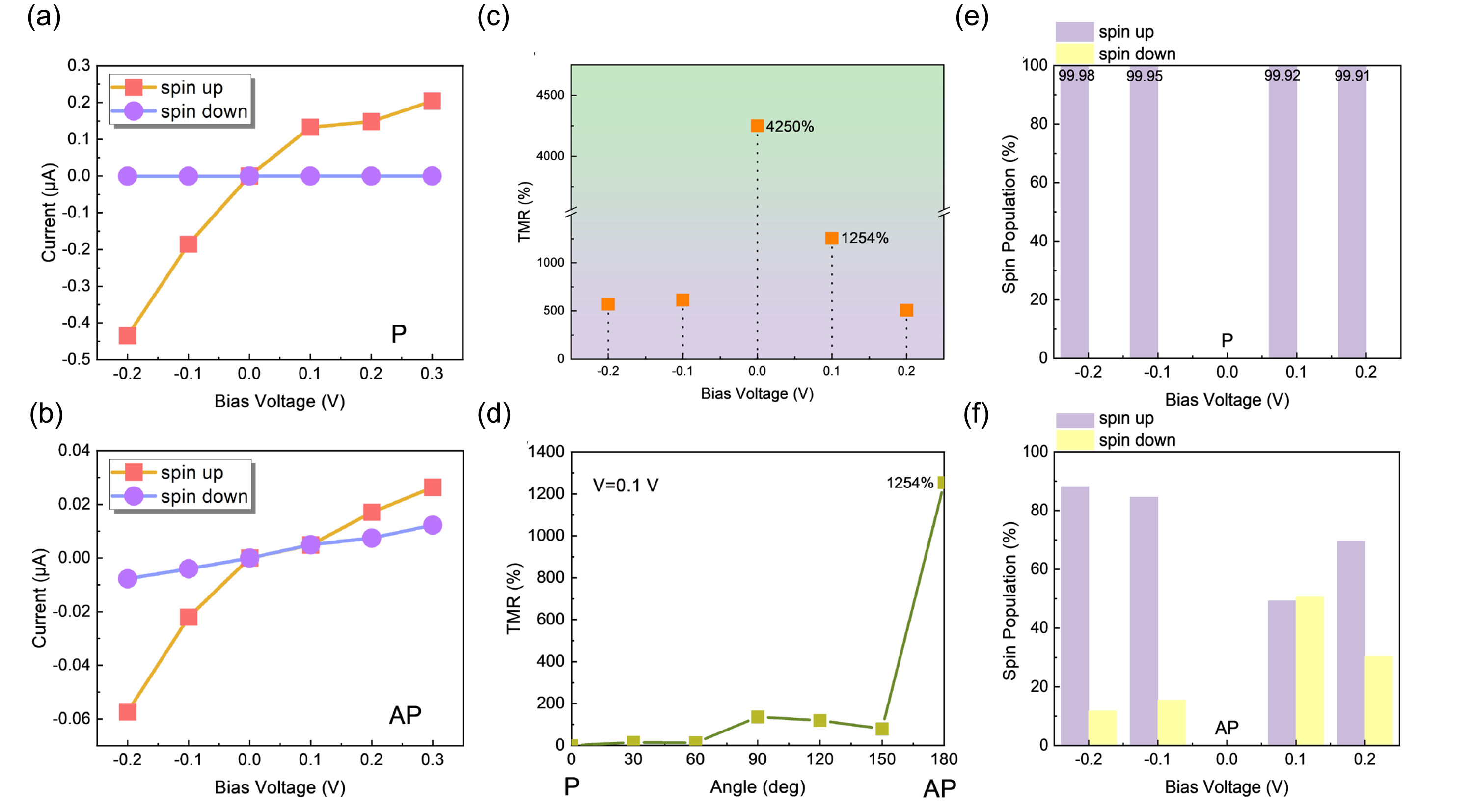}
\caption{The spin-resolved current of spin up and spin down for (a) P and (b) AP states. (c) Bias dependence of TMR ratio and (d) Angle $\theta$, between two magnetizations on two monolayers of LaBr$_2$, dependence of non-collinear TMR with fixed bias of 0.1V. Spin polarization of (e) the P state and (f) AP state.}
\label{fig:4}
\end{figure*}

The TMR of the such vertical MTJ at the equilibrium state (zero bias voltage) is first calculated. At the equilibrium state, the TMR is defined as $TMR = (T_P - T_{AP}) / T_{AP}\times100\%$, where $T_{P(AP)}$ represents the transmission coefficient at the Fermi level for the P and AP state, respectively. A large TMR of 4250\% is obtained. This is due to the transmission coefficient of the P state being over an order of magnitude larger than that of the AP state. A relatively larger $T_P$ and a smaller $T_{AP}$ contribute to a considerable TMR. To understand the origin of this high TMR effect, the $k_\parallel$-resolved transmission spectra are plotted in Fig. S3. The conductance is the integration of transmission over all $k_\parallel$ points in the 2D Brillouin zone. In the spin up of the P state, 
stronger peaks with $T_{PC\uparrow} = 0.16$ at the $\Gamma$ point results in the high conductance of the P state. 

Under the nonequilibrium state or bias, Fig. \ref{fig:4}(a-b) show the I-V curves for the spin up and spin down states in both the P and AP states. It can be seen that the current in the P state is an order of magnitude larger than that in the AP state, indicating good TMR performance. The current in the spin down channel for the P state is unchanged in an "off" state, as shown in Fig. \ref{fig:4}(a). In contrast, for the AP magnetic state, the current in the spin down channel generally increases with the bias voltage. Moreover, both the spin up and spin down currents exhibit an "on" state, which is different from the behavior observed in the P magnetic state. The calculated TMR is shown in Fig. \ref{fig:4}(c), which is defined as TMR = $\frac{{I_P - I_{AP}}}{{I_{AP}}} \times 100\%$. Here, $I_P$ and $I_{AP}$ represent the currents for the P and AP magnetic states, respectively. Notably, TMR is inversely proportional to the current for the AP state. Therefore, as the bias increases or decreases, the TMR decreases due to the enhancement of current for the AP state. Fig. \ref{fig:4}(d) shows the angle dependence of TMR under 0.1 V bias, which is defined as TMR($\theta$) = $\frac{{I(\theta) - I(0)}}{{I(0)}}$, where $I(0)$ represents the current for the P state with the magnetization of the bilayer LaBr$_2$ aligned in parallel, and $I(\theta)$ represents the current for a polar angle $\theta$ between two magnetization directions, and $I(\theta=180^\circ)$ corresponds to the AP state. It is found that during the switching process from the P state ($\theta=0^\circ$) to the AP state ($\theta=180^\circ$), the TMR gradually increases and finally reaches the maximum value. Besides TMR, the spin population, denoted as SP$\alpha$ = $\frac{{I\sigma}}{{I_{\text{Total}}}} \times 100\%$ ($\sigma = \text{up, down}$), is calculated, as illustrated in Fig.\ref{fig:4}(e-f). The results clearly demonstrate that, in comparison to the AP state, the spin population of the P state can achieve and sustain nearly 100\%, indicating an ideal spin-filtering characteristic that selectively permits the passage of spin-up current only, which is supported by the transmission spectra (Fig. S5).

In Fig. \ref{fig:4}(a-b), we note that the spin down current is not zero in the AP configuration, leading to low spin polarization. This problem can be addressed by constructing a double spin-filter tunnel junction\cite{miao2009magnetoresistance}, inserting an insulating monolayer $h$-BN between the bilayer LaBr$_2$. Unlike conventional MTJs, the two insulating magnetic LaBr$_2$ selectively filter one of the two electron spin directions \cite{moodera2007phenomena}, resulting in perfect spin polarization under positive bias as shown in Fig. S6(b-c). We also calculate the rectification ratios of the vdW LaBr$_2$/LaBr$_2$/WTe$_2$ and LaBr$_2$/BN/LaBr$_2$/WTe$_2$ MTJs under different bias voltages (Table S1). A clear comparison reveals that the inclusion of $h$-BN leads to improved rectification ratios, reaching up to 78.56 \%.

In conclusion, we have designed all-van der Waals AFM-LaBr$_2$/WTe$_2$ heterostructures with exceptional data write and read capabilities. Compared to devices made of 3D ferromagnetic materials, our two-sublattice macrospin simulations show that the use of 2D material LaBr$_2$ allows for faster switching at comparable or even lower current densities. The remarkable out-of-plane spin Hall conductivity of WTe$_2$ and delocalized nature of electride LaBr$_2$ enable such fast and efficient spin-orbit torque switching of magnetization. Additionally, a high TMR ratio of up to 4250\% and a high spin polarization of close to 100\% in the proposed system are achieved. The underlying microscopic physics responsible for this intriguing read performance have been analyzed through the examination of transmission spectra. The excellent performance highlights the tremendous potential of vdW PMA AFM LaBr$_2$ in advancing spintronic device applications. Specifically, it opens avenues for the design of ultra-low power, ultra-fast data operation, ultra-high density, and thermally stable memories.

\section*{acknowledgement}
The authors thank the support from the National Natural Science Foundation of China (Grant No.12104301, No. 51671114 and No. U1806219), MOE Singapore (MOE2019-T2-2-030, A-0005241-01-00 and A-8001194-00-00). This work is also supported by the Special Funding in the Project of the Taishan Scholar Construction Engineering.

\bibliographystyle{apsrev4-1}
\bibliography{v0/Bibliography.bib}

\end{document}